\documentclass{article}
\usepackage{graphicx}
\begin{document}
\title{Schwarzschild black holes and propagation                                
of electromagnetic  and gravitational waves.
\thanks{ Presented by E. Malec at the XXV International School of Theoretical Physics,
                          Particles and Astrophysics - Standard Models
                           and Beyond
                          10 - 16 September 2001 Ustro\'n, Poland}                          			  
}
\author{
Janusz Karkowski, Edward Malec
\\
Institute of Physics, Jagiellonian University,\\
30-059 Krak\'ow, Reymonta 4, Poland
\\                                                                            
\\
Zdobys\l aw \'Swierczy\'nski
\\
Institute  of Physics,  Pedagogical University,\\                       
30-  Krak\'ow,  Podchor\c a\.zych 1, Poland
}                                   
\maketitle
\begin{abstract}
 
Disturbing of a spacetime geometry may result in the appearance of an  
 oscillating  and   damped   radiation       - the so-called
quasinormal modes. Their periods  of  oscillations 
  and   damping coefficients carry unique information about the mass
and  the  angular momentum, that would allow one to identify the
source of the gravitational field. In this talk we
present recent   bounds on the diffused  energy, applicable to the 
Schwarzschild spacetime, that give also rough estimates of   the energy 
of excited quasinormal modes.
\end{abstract} 
 
\clearpage

\section{Introduction}

Quasinormal modes are scattering-type solutions of the Schr\"odinger
 equation which satisfy  a quite peculiar boundary condition - that 
at both "ends"  the waves are purely outgoing. 
They are being studied in the context 
of general relativity (the  Schr\"odinger equation emerges there through
the standard separation of the time dependence in a wave equation)
in the course of the last thirty years (\cite{Chandra}-\cite{Nollert}).
 Much is known about their eigenvalues and 
their temporal evolution in the  case of Schwarzschild   black holes and
neutron stars. An exhaustive   review on that topic is \cite{Bernd};
see also \cite{Nollert}.
 An observer located at a fixed   space 
position would find   that
   quasinormal modes oscillate in time and their
amplitude exponentially decays. The period  of  oscillations 
  and   damping coefficients carry unique information about the mass
and  the  angular momentum.  

Perturbation  of a  spherically symmetric spacetime geometry
\cite{exception}  
 can be described - as far as the backreaction
can be neglected - by a linear wave equation (\cite{Wheeler}
 - \cite{Zerilli} ).
The interesting fact is that a (compact support) 
perturbation  of  a spacetime geometry may result 
in the appearance of an  outgoing radiation that coincides in 
a bounded  region of spacetime  with a linear combination of  
quasinormal modes.
 At some intermediate time - later on the
so-called tail  term dominates -  the perturbation
is dominated by the fundamental mode, since  the latter is   damped
weaker than the other quasinormal  modes. The oscillation periods
and damping coefficients do not depend on perturbations; this feature
can be used in order  to identify the source of the gravitational
 field \cite{Damour}. 

It is of interest to know how much energy can be carried
by quasinormal modes. Their energy can be estimated by the so-called
diffused energy - the energy loss that is due to the backscattering.
 In this paper we present recent results in this direction. 
The order of the   rest of this paper is as follows: 
 
i) Space-time curvature and two patterns of propagation of massless  
fields; 

ii) Vibrations of a spacetime - an example;  
 
iii)  Recent results on the energy diffusion  in the Schwarzschild
 spacetime;

iv) Dependence of backscatter on   the frequency of waves;

v)  Discussion;

vi)  Lessons from numerics.

\section{What is backscattering? }
 
Let  an outgoing
null cone 
$\tilde \Gamma_{a}$ originate from $(a,0)$. 
Assume that  a flash of  radiation 
is  initially  purely  outgoing \cite{tricky}  and that its  support is 
contained in  an annular region $(a, b)$, $b\le \infty $.   
Then, depending on whether or not the spacetime is curved, 
the following can be observed. 
 
i) In the flat Minkowski spacetime a wave   remains purely 
 outgoing.   No radiation can be found in the interior of the cone (Fig. 1).
In this case no backscatter (\cite{Hadamard}, \cite{Hadamard1})
 occurs (in the Hadamard's  terminology:
the type B Huyghens principle holds true). 
%

\begin{figure}[h]
\centerline{\includegraphics[width=6cm]{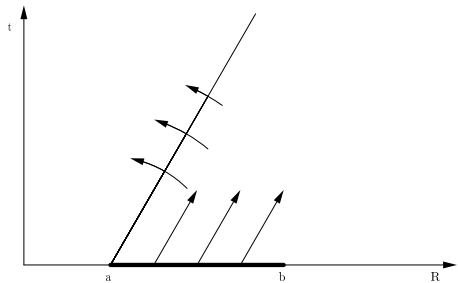}}
\caption{Solid line represents an outgoing null cone. Arrows show the 
direction of the radiation.}
\end{figure}

ii)   A wave  backscatters in a curved, e.g.  Schwarzschild, 
spacetime. Some energy, denoted later as $\delta E_a$, 
 diffuses   inward through   $\tilde \Gamma_{a}$
and is lost from the main stream (Fig 2).

\begin{figure}[h]
\centerline{\includegraphics[width=6cm]{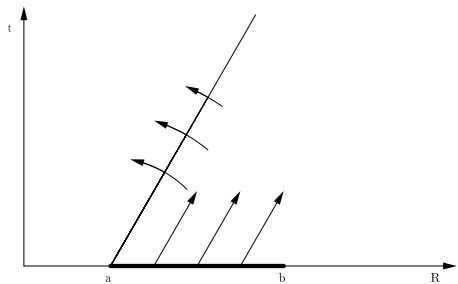}}
\caption{Solid curve represents an outgoing null cone. 
Arrows show the  direction of the radiation - there appears an
ingoing component.}
\end{figure}
 
%
One can shortly state that  waves propagate in  a curved
spacetime like  electromagnetic waves 
 in  a medium  with a varying refraction index.  
A fraction of the  radiation  scatters  
off the curvature of the  geometry  and   
a part of the initial energy never reaches infinity. 
  
\section{Damped oscillations} 
 
Figure 3 \cite{courtesy} shows a picture that is typical in the case
of perturbations that have compact support. An initial perturbation
is depicted in Fig. 3(a). Some time after the initial pulse runs  
through the observer, he (or she) can observe
that  an oscillatory (single frequency) radiation 
dominates.  As pointed out before, the period and damping coefficients
are independent on the perturbation.

There are  two interconnected  problems that can 
made difficult the identification
of the dominant mode. First, in the
 asymptotic zone, $t>>2m$, the tail terms 
dominate \cite{Bernd}, since they decay as 
some power of $1/t$. Second,  the amplitude of the
oscillations quickly decreases, as exemplified by 
Fig. 3(b).

\begin{figure}[h]
\centerline{\includegraphics[width=8cm]{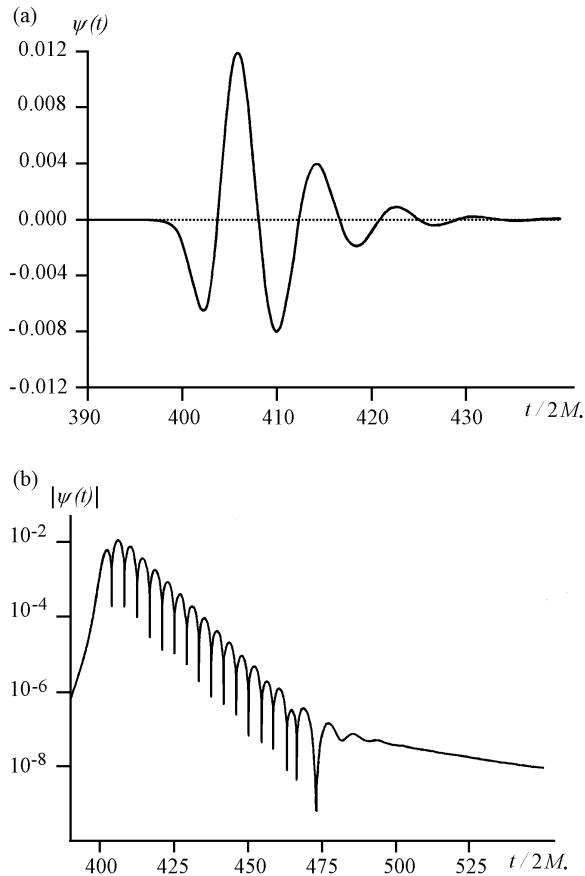}}
\caption{Time evolution of the Regge-Wheeler function for $L=2$.
The 'observer' is located at $r_{*}=800*m$ }
\label{FIG3}
\end{figure}

\section{  Quantitative estimates of the backscattered energy.} 

We specialize, beginning from this section, to spherically symmetric
spacetimes.  The backscattered (diffused) energy will be 
found from the energy conservation and some potential estimates. Let us
point out that our intention is to derive  a bound from above on the
fraction of the backscattered energy; this
bound should be  independent of the
details of the initial pulse of the outgoing radiation.
The only information that is required, is the mass $m$
of the central object  and the initial position  $a$ of the inner
 boundary of the outgoing radiation. (In the case of nonspherical
spacetimes one presumably would need also the information about the
total angular momentum.)

The idea that it is possible to bound the backscattered 
energy entirely in terms of  initial energy and initial position
was first put on trial  in the example of the massless 
scalar field (\cite{MENOM} - \cite {ME}; a recent work is
\cite{Roszkowski}).
The Theorem formulated below summarizes results that have been obtained 
while implementing a programme that was formulated in the
case of electromagnetic radiation in (\cite{malec2000}, 
\cite{meprl}) and then modified in  \cite{Gerhard} and \cite{kark}.  
While the analytic part concerning the electromagnetic radiation
is fairly satisfactory (any further progress would most likely involve
advanced numerical methods), the work on the gravitational 
radiation is not finished.    Preliminary results on the polar modes
are given in \cite{Gerhard} while the investigation of axial modes is
in progress.

 {\bf   Theorem.} 

 Assume a flash of dipole electromagnetic radiation (quadrupole 
 gravitational  radiation - axial or polar)
that is initially purely   outgoing  and that has  support in an annular
region $(a, b)$, $b\le \infty $.
 Let the initial energy be $E_a(0)$ and $\tilde m=m/a$.

Then the fraction of the diffused energy  $\delta E_a/E_a(0) $
satisfies the inequality 

\begin{center}
{ 
\frame{
\begin{minipage}{10cm}
\begin{displaymath}
{\delta E_a\over E_a(0)} \le  C(\tilde m) \Bigl( {2m\over a}\Bigr)^2.
\end{displaymath}
\vspace{0.03cm}
\end{minipage}
}
}
\end{center}

Here

i) (the case of electromagnetic radiation)
 for $a \ge 10{m \over 3}$, $0.3=C(0)\le C(\tilde m)\le C(10/3)\approx 1.7$  
\begin{eqnarray}
C(\tilde m)&=& {-1\over 10080((-1+2\tilde m)^4\tilde m^4)}\Biggl(
-2760\tilde m^5+828\tilde m^4+\nonumber \\
&& 44\tilde m^3+2352\tilde m^6+2016\tilde m^4\ln (1-2\tilde m)+
\nonumber \\
&&2688\tilde
m^6\ln (1-2\tilde m)-  
4032\tilde m^5\ln (1-2\tilde m)-\nonumber \\
&&360\tilde m^3\ln (1-2\tilde m)+
6\tilde m+36\tilde m^2\ln (1-2\tilde m) 
 -
 \nonumber \\
&&
 30\tilde m^2+3\ln (1-2\tilde m)-18\tilde m\ln (1-2\tilde m)\Biggr) ;
\nonumber \\
\end{eqnarray}
ii)  (axial gravitational  waves ) \cite{uwaga}
 $1.05=C(0)\le C(\tilde m)  = ? $  ;
 
iii)  (polar gravitational waves) \cite{uwaga}
 $55/2=C(0)\le C(\tilde m)=  ? $ 
 (conjecture: $C(0)<4$).

\vskip 1cm

{\bf Sketch of the  proof  }

\subsection{ Preliminaries.  }
 
A Schwarzschildean   line element reads (we neglect 
the backreaction effect):
\begin{equation}
 ds^2 = - (1-{2m\over R})dt^2 +
{1\over 1-{2m\over R}} dR^2 +
R^2 d\Omega^2~.
\end{equation}
The  radial terms of the multipole expansion  satisfy the
following (reduced) wave equations
 (\cite{Wheeler}, \cite{Zerilli}, \cite{Wheeler55}),

\begin{equation}
(-\partial_0^2 + \partial_{r^*}^2)\Psi_l = V(R)\Psi_l.
\label{4.1}
\end{equation}  
In the the  case of  lowest multipoles we have

i)   $l=1$  and  $V(R)=(1-{2m\over R}){ 2\over R^2}$ 
(dipole term, electromagnetism);  

ii) 
$l=2$ and   $V(R)=(1-{2m\over R}) { 6\over R^2}\Biggl( 1-{m\over R}
\Biggr) $ 
(quadrupole term, axial gravitational waves);

iii) $l=2$ and   $V(R)=(1-{2m\over R}) { 6\over R^2}\Biggl( 1-{2m\over R}
+{21m^2(1+{m\over R})\over 4 R^2(1+3m/(2R))^2}\Biggr) $ 
(quadrupole term, polar gravitational waves).

  In what follows we shall deal only with  the electromagnetic case;
the other two cases can be  treated   similarly.
Although  the calculations
become more complex, basic scheme is the same. 
 
The most general solution  of the dipole electromagnetic
radiation  in the Minkowski spacetime is given by 

\begin{equation} 
\partial_0\Bigl( f ( R-t) - g(R+t)\Bigr)
+{ f(R-t)+ g( R +t)\over R},
\end{equation}
where the $f$-related part describes the outgoing radiation while $g$ is
responsible for the ingoing wave. For any initial data, one can uniquely 
determine $f$ and $g$ and in this way specify the outgoing and ingoing 
initial pulses.

We invoke to the above decomposition in order to specify what is 
meant by in- or out- directed waves in the Schwarzschild spacetime.
Define  the Regge-Wheeler variable 
$   r^*(R) \equiv R+2m\ln ({R\over 2m}-1) $.  Then having initial data,
one can construct functions  $f$ and $g$ in a similar way as in the
Minkowski spacetime.
 If from the
\cite{uwaga}
 construction follows that $g=0$, 
then we will say that the radiation is purely outgoing. In such a case
it is useful to define
\begin{equation}
\tilde \Psi (R, t)\equiv -\partial_{r^*}f( r^*-t)
+{ f( r^*-t)\over R(r^*)},
\end{equation}
  and to seek a solution of the dipole wave equation,
$\Psi (r^*,t)$, having the following form   
\begin{equation}
\Psi =\tilde \Psi  +\delta .
\label{4.2}
\end{equation}
We would like to point out that at $t=0$
 the function $\delta$ vanishes up to the first time derivative
 \cite{malec2000}, $\delta = 0 $ and $\partial_0\delta =0 $.  
One finds that the evolution equation reads
\begin{equation}
(-\partial_0^2 + \partial_{r^*}^2)\delta = (1-{2m\over R})
\Biggl[ { 2\over R^2}
\delta + {6mf\over R^{4}}  \Biggr] .
\label{4.3}
\end{equation}
 
\vskip 1cm
\subsection{ Main steps of the proof.}

The energy of the electromagnetic field that is contained in the annulus
$(R(t), \infty )$ reads 
$$
E_{R(t)} =2\pi  \int_{R(t)}^{\infty }dr
\Biggl(  {(\partial_0\Psi )^2\over 1-{2m\over r}} + (1-{2m\over r})
(\partial_r\Psi )^2+{2(\Psi )^2\over r^2}\Biggr) 
$$ 

\subsubsection{ Bound  on  $f$ in terms of   the initial energy.}

 {\bf Lemma 1.} Define  $\eta_R\equiv 1-{2m\over R}$, $\tilde m\equiv m/a$ and  
$y\equiv R/a$.  Let $f(a)=0$, and $a\ge 10m/3$.  Then
\begin{equation}
|{f^2(R,0)\over R^2}| \le {E_a(0)\over 8\pi }a\eta^2(R)F(\tilde m, y),
\end{equation}
where 
\begin{eqnarray}
&&F(\tilde m, y)\equiv y-1+{16\tilde m^4\over 3(-y+2\tilde m)^3}
-{16\tilde m^4\over 3(-1+2\tilde m)^3}-\nonumber \\
&&{16\tilde m^3\over (-y+2\tilde m)^2}+
{16\tilde m^4\over (-1+2\tilde m)^2}+ {24\tilde m^2\over (-y+2\tilde m)}-
\nonumber\\
&&
{24\tilde m^2\over (-1+2\tilde m)}+8\tilde m\ln {y-2\tilde m\over 1-2\tilde m}.
\nonumber\\
\end{eqnarray}

This is essentially a Sobolev-type estimate. For the proof see \cite{kark}.

  \subsubsection{Estimate of  an "energy"  of $\delta $}

This "energy" denoted as $H$  is not conserved - 
there is a volume-dependent term in
 the integral form of this "energy"  evolution law.
 It appears to be, however,
a useful quantity.   $H$ is defined   by
\begin{equation}
H(R,t)\equiv \int_R^{\infty }dr \Bigl( {(\partial_0\delta )^2\over \eta_r}+
\eta_r(\partial_r\delta )^2+{2\delta^2\over r^2}\Bigr) ;
\label{4.4a}
\end{equation}
this integral is done on a fixed Cauchy hypersurface $t=const$.
One can prove

{\bf Lemma 2.}   Let  the support of initial data  be $(a, b)$, $a\ge 10m/3$, 
$b\le \infty $.  Then
\vskip 1cm
\begin{equation}  
H(a_t,t)\le 36m^2\Biggl[ \int_0^tds 
\Biggl( \int_{a_s}^{\infty }dr {f^2\eta_r\over r^8}\Biggr)^{1/2}\Biggr]^2 , 
 \end{equation}
where the $t$-integration follows along $\tilde \Gamma_{a, (a_t, t)}$ while
the $r-$integration is done on a fixed Cauchy slice.

 The crucial point in the proof of Lemma 2 is that 
\begin{equation}  
H(a_t,t)=-{\delta E_a(t)\over 2\pi } -12 \int_0^tds\int_{a_s}^{\infty }dr 
\partial_0\delta {f\over r^4} ; 
 \end{equation}
dropping out the nonpositive $\delta E_a$ terms and using the Schwarz inequality 
yields the lemma \cite{kark}.    
        
 \subsubsection{The energy conservation and bounding of the
diffused energy  (the energy ``loss'')}

Define $\tilde \Gamma_{R_0, (R,t)}$ -  a segment of an outgoing null geodesic that
connects $(R_0, t=0)$ with $(R,t)$.
Further, let us introduce
\begin{equation}                  
h_-(R,t) ={1\over 1-{2m \over R}}(\partial_0+\partial_{r^*})\delta ;
\label{4.4}
\end{equation}
this corresponds to this component of the "reduced" strength field  tensor
that is directed inward. 
\vskip 1.cm

 The rate  of the energy change along $\tilde \Gamma_{a}$
 is given by 
\begin{eqnarray}
&&(\partial_0+\partial_{r^*})E_{{a}}= \nonumber\\
&& -2\pi (1-{2m\over R})\Biggl[ (1-{2m\over R})\Bigl( {\partial_0\Psi \over
1-{2m\over R}}
 +\partial_R\Psi \Bigr)^2  +{ 2\over R^2}\Psi^2 \Biggl] =
\nonumber\\ &&
-2\pi (1-{2m\over R})\Biggl[ (1-{2m\over R}) \Biggl( h_--
 { f\over R^{2}}\Biggr)^2
  +{ 2\over R^2}\Bigl( \tilde \Psi+\delta\Bigr)^2 \Biggl] ;
\label{4.5}
\end{eqnarray}

but  $f =\tilde \Psi =0$ while their first derivatives   
vanish on   $\tilde \Gamma_{a}$.
The integration of (\ref{4.5}) along $\tilde \Gamma_{a}$ gives 
the energy that diffused through the segment $\tilde \Gamma_{a, a(t)}$,
\begin{eqnarray}
&& \delta E_a(t)\equiv  E_{a}(0)- E_a(t)=
 \nonumber\\ &&
2\pi \int_{a}^{a(t) } dr
  \Biggl[ (1-{2m\over r}) h^2_-
  +{ 2 \delta^2\over r^2} \Biggl] . 
  \label{4.5a}
\end{eqnarray}
>From the $H$-conservation law and   Lemma 2 one gets

\begin{equation}  
{\delta E_a\over 2\pi }\le 36 m^2\Biggl[
\int_0^tds\Biggl( \int_{a_s}^{\infty }dr {f^2\eta_r\over
 r^8}\Biggr)^{1/2}\Biggr]^2 ; 
 \end{equation}
here $\delta E_a\equiv \lim_{t\rightarrow \infty }\delta E_a(t)$.
 \vskip 1cm
The "electromagnetic" part
of the main Theorem follows from the preceding bounds.

    \section{Dependence of the  backscatter on the  frequency of waves}

The backscattering depends on the relative width of support.
That is well known from the numerical analysis of Vishveshwara
 \cite{Vishveshwara}),
but the first - up to our knowledge -   proof appeared in  \cite{meprl}.
Below we sketch the main results.
When   support of the
initial radiation is very narrow, $\kappa =(b-a)/a<<1$, then
\vskip 1cm
$$
{\delta E_a\over E_a(0)} \le C\Bigl( {2m\over a}\Bigr)^2\kappa
   ,
$$
In the   limit $ \kappa \rightarrow 0$   
 ${\delta E_a\over E_a(0)}   \rightarrow 0$;
  the backreaction is negligible. We would
like to point that this argument works for all $a>2m$, in contrast to
the main Theorem.
 
The question  arises: how do we interpret this? 
The {\it similarity theorem} \cite{Bracewell}
 of Fourier transform theory
states that compression of support of a function corresponds to expansion  of
the frequency scale.  It means  that
\begin{equation}
\lim_{\kappa \rightarrow 0}{E( \omega \le \Omega_c)\over E_a(0)}
\rightarrow 0;
\end{equation}
above  $E( \omega \le \Omega_c)$  is the electromagnetic energy of modes
with frequencies smaller than a fixed frequency, $\omega \le \Omega_c$.
Thus we can conclude  that the    high frequency radiation is
 essentially unhindered by the  backscattering     
while long waves can be backscattered.

\section{Discussion}

The diffused energy  $\delta E_a$ bounds from above the sum 
$\delta E_{qmt}+\delta E_f$ where $\delta E_{qmt}$ is the  energy carried
 by  the quasinormal modes and the tail, and 
$\delta E_f$ is the energy of that radiation that 
falls to a black hole or hits the surface of a star. 
Below we present several data that estimate from above the total
fraction of the backscattered energy. A more
detailed calculation would show that the  last contribution, $\delta E_f$,
 dominates, so that we expect that  $\delta E_{qmt}/E_a(0)$
 can be a small fraction of the number given in forthcoming examples.
 
In the first group of examples we consider the case, 
when the initial pulse is  close to the Schwarzschild radius $R=2m$;
   
 i) $a=10m/3$ (e. g., a surface of a  supercompact neutron star  ): 
 ${\delta E_a \over E_a(0)}<0.5$;

ii) $a=4m$ (e. g., a  supercompact neutron star  ):   $ {\delta E_a \over E_a(0)}< 0.3$; 
 
iii)   $a=5m$ (a standard neutron star): ${\delta E_a \over E_a(0)}< 0.13$.

It is instructive   to notice that this effect is very weak in the case
of other  astronomical objects.   For the  Sun for instance : 
  ${\delta E_a\over E_a(0)} \approx
 10^{-13}$, while for  white dwarves: ${\delta E_a\over E_a(0)}<10^{-7}$.
Our estimates allow  one also to answer
the   question  that  was raised in the literature  (see, e.g., \cite{Bonnor}),  
 how strongly gravitational field  of a galactic interior
 can damp the outgoing radiation? 
One can infer from the preceding examples that
the effect may matter only, if there is a black hole in the interior
of a galaxy.

\section{What can be learned from numerics?}
 
We summarize shortly the main conclusions that can be obtained 
through a numerical analysis. 
 
i)  If a radiating source is close to a horizon, 
then the damping can be quite strong. We found,
in accordance with expectations, that  the
 backscatter is strong deep inside the photon sphere. There 
are known   examples  in which from 20 percent 
 (electromagnetism) to 49  percent (gravitational 
polar modes) of the initial energy gets 
diffused \cite{kark}; that  
should go to 100 \% if a very compact source is
infinitely close to the event  horizon.

ii)   The efect depends on the width of
a"renormalized" ($(b-a)/\eta_a$) support of initial energy. The backscatter 
 is strongest, when the 
width is of the order of $a$ - comparable to the areal distance from the center.

iii)   The present analytic estimates are  not exact.  They are
  expected to yield,  in combination with
appropriate numerical methods, more satisfactory results.

{\bf Acknowledgements.}  
 
This work has been suported in part  by
the KBN grant 2 PO3B 010 16.  One of the authors (EM) thanks Gerhard
Sch\"afer for many discussions.


\begin{thebibliography}{100}
\bibitem{exception} In the case of stars, one of the boundary
conditions is replaced by a condition at the center and/or 
at the surface of a star \cite{Bernd}.
\bibitem{tricky} The notion of "outgoing waves" is somewhat tricky 
in curved spacetimes. See a discussion in Sec. IV. B.
\bibitem{courtesy} This figure is taken from \cite{Nollert}, with
the permission of Peter-Hans Nollert and the Institute of Physics.
\bibitem{uwaga} The upper range of the validity of the formula is still
uncertain. If $a>4m$ then we expect to obtain a nontrivial estimate.
\end{thebibliography}
\end{document}